\documentclass[12pt]{article}
\usepackage{amsmath, amssymb,graphicx}

\usepackage{amssymb}
\usepackage{graphics}
\begin{document}
\baselineskip 18pt
\newcommand{\Tr}{\mbox{Tr\,}}
\newcommand{\beq}{\begin{equation}}
\newcommand{\eeq}{\end{equation}}
\newcommand{\bea}{\begin{eqnarray}}
\newcommand{\eea}[1]{\label{#1}\end{eqnarray}}
\renewcommand{\Re}{\mbox{Re}\,}
\renewcommand{\Im}{\mbox{Im}\,}

\def\N{{\cal N}}
\def\one{\hbox{1\kern-.8mm l}}

\thispagestyle{empty}
\renewcommand{\thefootnote}{\fnsymbol{footnote}}

$\left. \right.$ \hfill MPP-2014-573
 \vspace{1cm}

\begin{center} \noindent {\Large \bf
Meson spectra of asymptotically free  gauge \medskip

theories from holography}
\end{center}

\bigskip\bigskip\bigskip

\begin{center}  \normalsize {\bf Johanna Erdmenger$^1$, Nick Evans$^2$  \& Marc Scott$^2$ }

\bigskip
\bigskip\bigskip

{ \it $^1$ Max-Planck-Institut f\"ur Physik (Werner-Heisenberg-Institut),

F\"ohringer Ring 6, D-80805 Munich, Germany

$^2$ STAG Research Centre and  Physics \& Astronomy,

 University of
Southampton, Southampton, SO17 1BJ, UK.} \bigskip

{\it jke@mpp.mpg.de, evans@soton.ac.uk, m.scott@soton.ac.uk} \end{center}
\bigskip

\bigskip\bigskip

\renewcommand{\thefootnote}{\arabic{footnote}}

\centerline{\bf \small Abstract}
\medskip

{\small \noindent Using holography, we study the low-lying mesonic spectrum of a range of asymptotically free gauge theories. First we revisit a simple top-down holographic model of QCD-like dynamics with predictions in the $M_\rho$ - $M_\pi$ plane. The meson masses in this model are in very good agreement with lattice gauge theory calculations in the quenched approximation. We show that the key ingredient for the meson mass predictions is the running of the anomalous dimension of the quark condensate $\gamma$. This provides an explanation for the agreement of holographic and quenched lattice gauge theory calculations.
 We then study the `Dynamic AdS/QCD model' in which the gauge theory dynamics is included by a choice for the running of $\gamma$. We use the naive two-loop perturbative running of the gauge coupling extrapolated to the non-perturbative regime to estimate the running of $\gamma$ across a number of theories. We consider models with quarks in the fundamental, adjoint, two-index symmetric and two-index anti-symmetric representations. We display predictions for $M_\rho, M_\pi, M_\sigma$ and the lightest glueball mass. Many of these 
theories, where the contribution to the running of $\gamma$ is dominated by the gluons, give  very similar spectra, which also match with lattice expectations for QCD. On the other hand, a significant difference between spectra in different holographic models is seen for theories where the quark content changes the gradient of the running of $\gamma$ around the scale at which chiral symmetry breaking 
is triggered at $\gamma \simeq 1$. For these walking theories we see an enhancement of the $\rho$ mass and a suppression of the $\sigma$ mass. Both phenomena are characteristic for walking behaviour in the physical meson masses.    }

\newpage
\section{Introduction}

Asymptotically free gauge theories are notoriously difficult to study since they run to strong coupling in the infra-red. Computing the bound state spectrum of theories such as QCD is therefore very hard. First-principle lattice calculations are possible but very numerically expensive. They are typically guided by the answers observed in nature. It is hard to explore the range of behaviour across the full space of asymptotically free theories. The holographic description of large $N_c$ ${\cal N}$=4 gauge theory \cite{Maldacena:1997re} has raised the prospect of a dual gravitational picture for these theories in which the spectrum might be computed in a purely classical theory.  Top-down attempts \cite{Karch:2002sh, ERev, Babington:2003vm}  to rigorously find a gravity dual originating from ten-dimensional string theory are complicated by the need to find a brane construction that decouples all unwanted super-partners, and also by the challenge of finding the appropriate gravitational background for embedding those branes. In any case when the gauge theory is weakly coupled, such as in the ultra-violet, the gravitational theory will itself become strongly coupled. Bottom-up holographic modelling \cite{Erlich:2005qh}  has taken broad brush stroke lessons from the AdS/CFT correspondence and attempted to model the mesonic and glueball degrees of freedom. Basic AdS/QCD models appear to work reasonably well, even at the quantitative 10$\%$ level or better, but are not systematically improvable. This is due to the fact that in principle, very many operators and higher dimension couplings can be important for the vacuum and bound state structure. Both top-down and bottom-up models have therefore struggled to encode the particular dynamics of a specific theory with, for example, a definite value of $N_c$ or the number of quark flavours $N_f$.

Recently there have been new attempts to construct holographic models \cite{others,JK,Alho:2013dka} that address these issues and provide insight into why  some top-down models give good descriptions of the QCD spectrum \cite{Alvares:2012kr}. Here we will push these insights further with two goals: First, we  provide further support for the success of an existing top-down model. Second, we present bottom-up models for a large range of different gauge theories. One key observation was highlighted already in \cite{JK}, where it was noted in particular that the quark condensate that characterizes the vacuum in all such models is described holographically by a scalar in an AdS-like space. The scalar becomes unstable to acquiring a vev, corresponding to a quark condensate, when its mass violates the Breitenlohner-Freedman (BF) bound \cite{Breitenlohner:1982jf}. This mass bound is given by $m^2=-4$ in AdS$_5$, for instance. The AdS mass of the scalar in turn is mapped by the AdS/CFT dictionary \cite{Maldacena:1997re}  to the dimension of the gauge theory operator, $m^2 = \Delta(\Delta-4)$. This implies that in a dual to a QCD-like theory,  the operator $\bar{q}q$ operator has dimension three in the UV, and is described by a scalar with $m^2=-3$. To reach
the BF bound, beyond which there is an instability leading to condensation to a new ground state and chiral symmetry breaking, the dimension of  $\bar{q}q$ must have become $\Delta=2$, which corresponds to an  anomalous dimension of $\gamma=1$. Top-down holographic models of QCD-like theories that use probe branes display the importance of these ideas: as described in \cite{Alvares:2012kr}, the running of the coupling or factors deviating from AdS in the background metric enter into the Dirac-Born-Infeld (DBI) action of the probe that describes the quark and meson physics. If the DBI action is linearized,  it leads to an action for a scalar in AdS, dual to the operator describing $\bar{q}q$,  with a running mass squared. This effective running mass is generated by the metric and forms of the background geometry which enter the DBI action. In fact, the dual geometry and running of the coupling enter into the effective AdS/QCD model only through the running of the anomalous dimension $\gamma$, i.e.~of the mass squared of the AdS scalar. 

Models that describe QCD reasonably have $\gamma=0$ in the UV (as occurs naturally in a theory that is supersymmetric in the UV), a long range over which $\gamma$ is small and a sudden rise to $\gamma$ greater than one. This running is broadly similar to that in QCD, where the logarithmic running keeps $\gamma$ small except around $\Lambda_{QCD}$ where it blows up rapidly. We will illustrate this here in the top-down Constable-Myers model \cite{Babington:2003vm, Constable:1999ch}, which has been higlighted in \cite{ERev} as providing a surprisingly good description relative to lattice data of the light spectrum.

These ideas were simplified in  \cite{Alho:2013dka}  where the `Dynamic AdS/QCD' model was proposed. The model is just the linearized DBI action of the D3/probe-D7 system, but with an arbitrary running for $\gamma$. To describe any particular gauge theory then requires a  guess as to the form of that running. A naive but still sensible guess is provided by the perturbative running of the QCD coupling to two loops. For $N_f$ just below $11 N_c/2$ where asymptotic freedom is gained, the two-loop running displays a Banks-Zaks fixed point \cite{Caswell:1974gg}. As $N_f$ is decreased, the value of the coupling at the fixed point increases and the anomalous dimension $\gamma$ increases. At a point close to $N_f \simeq 4 N_c$ for fundamental quarks, the BF bound is tripped and chiral symmetry breaking sets in. Above that value of $N_f$, the model is in the regime referred to as the ``conformal window''  \cite{Ryttov:2007cx}. Using the standard AdS relations, the running can be translated into a radially dependent mass squared for the scalar describing the condensate. The model then makes predictions for the spectrum of the theory. Here we will concentrate on the $\rho$ meson, the pions, the $\sigma$ meson ( i.e.~the singlet $\bar{q}q$ bound state with vanishing quantum numbers, also identified with the $f_0$) and the lightest glueball. For the glueball, only qualitative statements are possible since the Dynamic AdS/QCD model concentrates on the quark sector. We will present our results in the style of ``Edinburgh'' plots \cite{Bowler:1985hr} used by lattice gauge theorists. These plots
display only physical observables, such as the mass of the $\rho$ as function of the pion mass,
in order to remove scheme-dependent quantities such as  the quark mass. 

In our section 2 we will revisit the Constable-Myers model of chiral symmetry breaking \cite{Babington:2003vm, Constable:1999ch} and extract, by linearizing the DBI action of a D7 brane in the geometry, the running anomalous dimension. We will show that in the critical range of radial coordinate, where $\gamma \simeq 1$, the running of the model is similar to quenched QCD. We compare the $M_\rho$ against $M_\pi$ behaviour with that of quenched lattice computations \cite{Bali:2013kia} and re-emphasise the surprising success of the model.  In the subsequent sections, we then turn to the Dynamic AdS/QCD model \cite{Alho:2013dka} which allows us to explore the space of gauge theories as a function of $N_c$, $N_f$ and in dependence on the representation of the quarks. Again we find the holographic models give good agreement with lattice data where it exists. In fact we find a relatively  weak dependence on $N_c$ and the quark representation. Significant deviations from QCD-like behaviour is seen for so-called ``walking'' gauge theories \cite{Holdom:1981rm}. These are theories whose running is governed by an IR fixed point, although this point is never reached due to chiral symmetry breaking. Moreover, these theories approach fixed points with $\gamma$ close to but above 1 since they exist for $N_f$ just below $N_f \simeq 4 N_c$. For these theories, the gradient of $\gamma$ as function of the running coupling  is small when $\gamma=1$. They are expected to have a quark condensate which is enhanced in the UV, which in turn tends to enhance the $\rho$ mass, enhance the $\pi$ mass to a lesser degree and to suppress the $\sigma$ mass \cite{Haba:2010hu,DelDebbio:2010ze}. The effective potential becomes very flat as the UV condensate is pushed out to high scales, leaving a pseudo-flat radial direction in the potential. 
We observe all of these phenomena in our model. 

These models remain only models though, since they cannot be brought closer to the true dynamics systematically, and depend on the naive guess for the running of $\gamma$.  Indeed in gauge theory beyond two loops the running parameters are gauge dependent but we hope working at lower order does correctly capture the dynamics of the gauge theory running. Nevertheless, the success in reproducing the lattice data, where such data exists, gives  hope that  the approach presented provides information about universal behaviour in these theories. In particular we hope that the trends we see as the edge of the conformal window is reached, should provide guidance to lattice simulations of those theories \cite{lattice}.

\bigskip
\section{A Top-Down Model}

An early holographic description of QCD \cite{Babington:2003vm} was provided by placing 
D7-brane probes in the dilaton flow geometry of Constable and Myers \cite{Constable:1999ch}. D3-D7 strings introduce quenched quark degrees of freedom. The Constable-Myers deformation of AdS$_5\times$S$^5$ is a very simple  description of a gauge theory with a running coupling that breaks the ${\cal N}=4$ supersymmetry completely. The non-trivial dilaton profile is dual to that running coupling and has an IR pole which is ill-understood. In practice the geometry describes a gravity dual of a soft wall since the singularity is repulsive to probe branes. The D7 probes bend away from the singularity and asymptotically the embedding describes a dynamically generated quark condensate at zero quark mass. In \cite{ERev} the light meson spectrum was computed.
Moreover, the $M_\rho$ versus $M_\pi^2$ plot was compared to quenched lattice data \cite{Bali:2013kia}. We update these computations in Figure 1. The fit is remarkably good. At the time this seemed very surprising since the gauge theory apparently lies close to infinitely strongly coupled ${\cal N} = 4$  gauge theory with all the associated super-partners and has no asymptotic freedom. In this section we  return to this model and analyze it in the spirit of \cite{Alvares:2012kr} to shed some light on the success at describing the QCD spectrum.

\begin{figure}[]
\centering
\includegraphics[width=16cm]{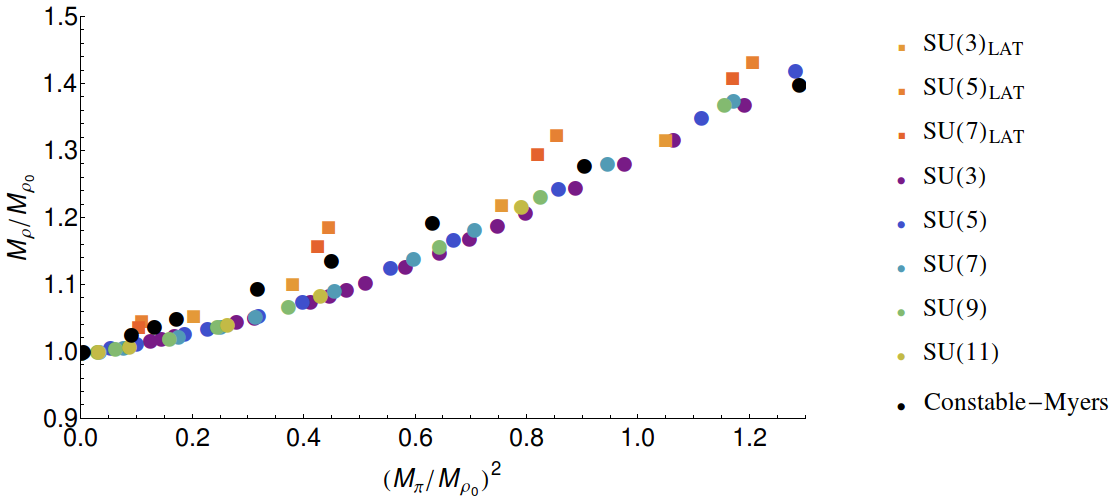}
\caption{Plots of $M_\rho$ against $M_\pi^2$ - in each case the points are normalized by $M_\rho$ at $M_\pi=0$ to set the non-perturbative scale $\Lambda$. As shown in the key, the plot shows the data for quenched lattice computations taken from \cite{Bali:2013kia} (and linearly fitted to find $M_\rho$ at $M_\pi=0$); the Constable Myers top down model; and the Dynamic AdS/QCD predictions.  }
\label{CMvslat}
\end{figure}

The gravity background of
Constable and Myers \cite{Constable:1999ch}  in {Einstein
  frame} has the geometry 
\begin{equation}  ds^2  =   H^{-1/2} \left( { w^4 + b^4 \over w^4-b^4}
\right)^{\delta/4} \, \sum\limits_{j=0}^{3} dx_{j}^2 
+ H^{1/2} \left( {w^4 + b^4 \over w^4-
b^4}\right)^{(2-\delta)/4} {w^4 - b^4 \over w^4 } \sum_{i=1}^6
dw_i^2 \,,
 \end{equation}
where $b$ is the scale of the geometry that determines the size of the
deformation ($\delta = R^4/(2 b^4)$ with $R$ the AdS radius) and
\vspace{-0.2cm}
\begin{align} H =  \left(  { w^4
+ b^4 \over w^4 - b^4}\right)^{\delta} - 1 \, ,   \qquad
w^2 = \sum\limits_{i=1}^{6} w_i{}^2
\, .
\end{align}
In this coordinate system, the dilaton and four-form are,
with $\Delta^2 + \delta^2 =10$,
\beq e^{2 \Phi} = e^{2 \Phi_0} \left( { w^4 + b^4 \over
w^4 - b^4} \right)^{\Delta}, \quad C_{(4)} = - {1 \over 4}
H^{-1} dt \wedge dx \wedge dy \wedge dz \,.
\eeq
This geometry returns to $AdS_5 \times S^5$ in the UV as may be seen
by explicitly expanding at large radial coordinate ~$w$.

To add quarks \cite{Babington:2003vm}
we will use an embedded probe D7-brane. The D7-brane will be embedded
with world-volume coordinates identified with $x_{0,1,2,3}$ and
$w_{1,2,3,4}$.  Transverse fluctuations will be parameterized by $w_5$
and $w_6$ (or $L$ and $\phi$ in polar coordinates) - it is convenient to define a coordinate $\rho$ such that
$\sum_{i=1}^4 dw_i^2 = d\rho^2 + \rho^2 d\Omega_3^2$ and the radial
coordinate is given by $w^2=\rho^2 + w_5{}^2 + w_6{}^2= \rho^2 + L^2$.

The Dirac-Born-Infeld action of the D7-brane probe in
the Constable-Myers background takes the form
\begin{equation} \label{CMD7} S_{D7} \,  =  \, - T_7 R^4
\int d^8 \xi~ \epsilon_3 ~  e^{ \phi} { \cal G}(\rho,L) 
 \Big( 1 + g^{ab} g_{LL} \partial_a L
\partial_b L  + g^{ab} g_{\phi\phi} \partial_a \phi
\partial_b \phi + 2 \pi \alpha' F^{ab}\Big)^{1/2}, 
\end{equation}
where
\begin{align} & {\cal G}
 = \rho^3 {( (\rho^2 + L^2)^2 + b^4) ( (\rho^2 + L^2)^2 - b^4) \over
(\rho^2 + L^2)^4} \,. \nonumber
\end{align}
Here we have rescaled $w$ and $b$ in units of $R$,  so that factors of $R$ only occur as an overall
factor on the embedding Lagrangian.

From these equations we derive the corresponding equation of
motion. We look for classical solutions of the form $
L(\rho),\, \phi =0$. 
Numerically we shoot from a regular boundary condition in the IR ($L'=0$) and
find solutions with the asymptotic behaviour $L \sim
m + c/\rho^2$.  These coefficients are then identified with the quark mass 
and condensate $\langle \bar{\psi} \psi \rangle$ respectively (formally $c$ is only the 
unique contribution to the condensate in zero mass limit \cite{Karch:2005ms}), in
agreement with the usual AdS/CFT dictionary obtained from the
asymptotic boundary behaviour.

Mesonic states are identified by looking at linearized fluctuations about the background embedding. Fluctuations in $\phi$ correspond to the pion and fluctuations in the world volume gauge field the $\rho$ meson. In each case one seeks solutions of the form $f(\rho)e^{ik.x}, k^2=-M^2$ with the mass states being picked out by the condition that  $f(\rho)$
is regular. 

Fig.\ref{CMvslat} shows the first example of the plots we will be producing in this paper - it shows the $\rho$ meson mass as a function of the pion mass squared. Note that in any given theory we must fix the strong coupling scale $\Lambda$. Here and throughout this paper we choose to do this by setting the $\rho$ mass at $M_\pi=0$ (ie when the quark mass is zero) the same in all theories, and we express all physical quantities in units of that fixed mass. The figure shows the results from the Constable-Myers model. We also display quenched lattice results for the plot in theories with gauge group SU(3), SU(5) and SU(7) - data taken from \cite{Bali:2013kia}. Note to place the lattice data on the plot we have taken the two data points at lowest $M_\pi$ and linearly extrapolated to find $M_\rho$ at $M_\pi=0$. This is naive and we will argue later that this maybe puts the points a little high in the plane. Conservatively we will use the spread of the lattice data across the different SU($N_c$) theories as reflective of the systematic errors in the lattice simulations.  The remarkable thing is the lack of dependence on $N_c$ in the lattice data and the match of the holographic model to the lattice data. The aim of this section is to identify why there is such a close match given the large deviations in the holographic dual that includes different adjoint particle content and UV behaviour.

Following \cite{Alvares:2012kr} we will argue that the key element for the quark physics in the top-down model is the running of the anomalous dimension $\gamma$ with the renormalization scale. We will show that this running is very similar to that in QCD, especially in the regime where $\gamma \simeq 1$ and where the BF bound-violating instability sets in that causes chiral symmetry breaking. To study this instability we will look at when the chirally symmetric $L=0$ embedding becomes unstable. We simply take our DBI action which up to a multiplicative constant  we may write as
\begin{equation}S_{D7}=\int d\rho\lambda(\rho,L)\rho^3\sqrt{1+(\partial_\rho L)^2},\end{equation}
where $\lambda(\rho,L)=\rho^{-3}e^\phi{\cal G}(\rho,L)$ and $r=\sqrt{\rho^2+L^2}$, and expand around $L=L'=0$ to quadratic order
\begin{equation}\begin{array}{rcl} \label{CMD7} S_{D7} \,&=&\int d\rho\,\, \rho^3\left(\left.\lambda\right|_{L=0}+\left.\frac{\partial\lambda}{\partial L^2}\right|_{L=0}L^2\right)\left(1+\frac{1}{2}\left(\partial_\rho L\right)^2\right) \, ,\\
&&\\
&=&\int d\rho\,\,\rho^3\left(\frac{1}{2}\left.\lambda\right|_{L=0}\left(\partial_\rho L\right)^2+\left.\partial_{L^2}\lambda\right|_{L=0}L^2   \right).              
\end{array}
\end{equation}

In order to ensure that the kinetic term in our Lagrangian is canonical, we perform a coordinate transformation on $\rho$,
\begin{equation}
\lambda(\rho)\rho^3\frac{\partial}{\partial\rho}\equiv\tilde{\rho}^3\frac{\partial}{\partial\tilde{\rho}},
\end{equation}
that is,
\begin{equation}
 \tilde{\rho}=\sqrt{\frac{1}{2}\frac{1}{\int_\rho^\infty\frac{1}{\lambda\rho^3}d\rho}}.
\end{equation}

We may rewrite our action in terms of the $\tilde{\rho}$-variable. Along with writing $L(\rho)=\tilde{\rho}\,\phi(\tilde{\rho})$, we obtain
\begin{equation}
 S_{D7}=\int d\tilde{\rho}~ \frac{1}{2}\tilde{\rho}^3\left(\tilde{\rho}^2(\partial_{\tilde{\rho}}\phi)^2+3\phi^2+\left. \lambda\frac{\partial\lambda}{\partial\rho}\right|_{L=0}\frac{\rho^5}{\tilde{\rho}^4}\phi^2  \right).
\end{equation}

The first two terms in the action describe a canonical $m^2=-3$ scalar in AdS$_5$, whereas the remaining term gives $\rho$-dependent mass to the scalar field in AdS$_5$. We find an overall mass squared
\begin{equation}
 m^2=-3-\delta m^2,\quad\quad \delta m^2\equiv-\left.\lambda\frac{\partial\lambda}{\partial\rho}\right|_{L=0}\frac{\rho^5}{\tilde{\rho}^4}.
\end{equation}
Using the standard scalar mass/operator dimension relation of the AdS/CFT dictionary, $m^2=\Delta(\Delta-4)$, but now assuming the mass dimension of the $q\bar{q}$-operator to be $3-\gamma$, where $\gamma$ is the running anomalous dimension of the gauge theory quark mass, we obtain the relation
\begin{equation}m^2=-3-2\gamma+\gamma^2.\end{equation}
Thus we associate $\delta m^2=-2\gamma+\gamma^2$, and are thus able to extract a running anomalous dimension in the Constable-Myers background.

The key point to note is that the only way that the background geometry and running dilaton enters into the equation for the embedding is through the running of $\gamma$. The background D7 embedding is then the key ingredient for the computation of linearized fluctuations that determine the mesonic masses. Effectively the origin of the running of $\gamma$ is lost - so questions about whether the background has too many superpartners of the gauge fields, or whether the running coupling is correctly that of QCD in the UV, and so forth become subsumed into simply asking whether $\gamma$ is close to that in QCD.  

\begin{figure}[]
\centering
\includegraphics[width=10.5cm]{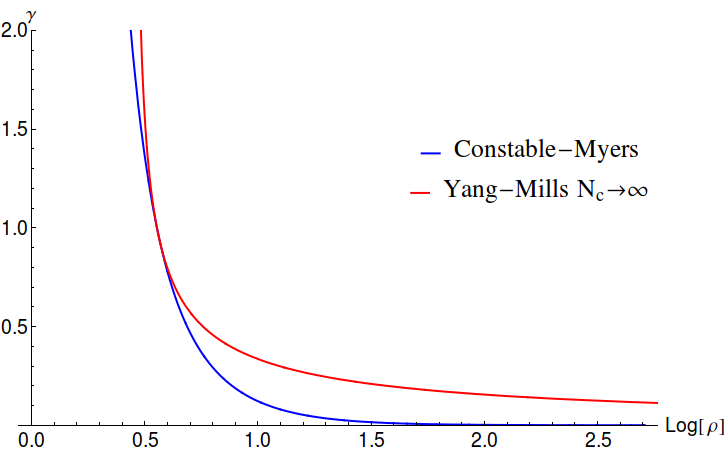}
\caption{A plot of the anomalous dimension $\gamma$ in the top-down Constable-Myers model. It is compared to  QCD by using  the one-loop perturbative result for the running coupling in large $N_c$ Yang-Mills theory ($\mu d \alpha/ d \mu = - 11 N_c \alpha^2 /6 \pi$) as input for calculating the anomalous dimension $\gamma$ ($\gamma = 3 N_c \alpha(\mu) /4 \pi$). We set the scale at which $\gamma=1$ to be equal in each case.}
\label{CMvsQCD}
\end{figure}

In Fig. \ref{CMvsQCD}, we plot the RG scale dependence of the anomalous dimension $\gamma$ extracted from the Constable-Myers model and the one loop running of large $N_c$ quenched QCD theory. We have matched the strong-coupling scale of the two theories by assuming that they each take the value $\gamma=1$ at the same scale.
Setting the AdS radius $R$ to one, we identify the RG scale and the radial coordinate by $\mu = \ln \, \rho$   (ie we are choosing to set this relation by matching to the physical RG scale).  
 This is the scale where chiral symmetry breaking is triggered, in the holographic model by the BF bound violation. From the figure it is immediately obvious that the scale dependence
of the anomalous dimension $\gamma$ is similar in both cases, and the gradient
of $\gamma$ is almost the same near $\gamma=1$. 
Deviations in the UV are present but are mild.  They occur in the regime where the BF bound is not violated in the holographic model. 

This close matching of the scale dependence of the anomalous dimension is, we believe, the reason for the success of the holographic model. It is worth pointing out that the reason that  the holographic description and QCD match in the UV is somewhat artificial. The UV of the Constable-Myers theory is infinitely strongly coupled ${\cal N}=4$ super Yang-Mills theory, yet the theory's large amount of supersymmetry preserves the perturbative dimension of the quark operator, i.e.~$\gamma=0$. In QCD, the UV result $\gamma=0$ simply follows from weak coupling. This coincidence has long been behind the successes of AdS/QCD models. 

Given that the key ingredient to describe the mesonic spectrum is simply the running of $\gamma$, it seems an obvious step to do away with the background construction of a geometry that mimics QCD, since there is no top-down holographic construction of real QCD, and to simply use the assumed form of $\gamma$ as an imput in the DBI action. This is essentially the starting point for the bottom-up model that we call `Dynamic AdS/QCD' \cite{Alho:2013dka}, which we will now move to studying.

\newpage
\section{Dynamic AdS/QCD}

Dynamic AdS/QCD was introduced in detail in \cite{Alho:2013dka}. The model maps onto the action of a probe D7 brane in an AdS geometry expanded to quadratic order \cite{Alvares:2012kr}. The anomalous dimension of the quark mass/condensate is encoded through a mass  term that depends on the radial AdS coordinate $\rho$.

The five-dimensional action of our effective holographic theory is
\begin{equation}
S   =   \int d^4x~ d \rho\, {\rm{Tr}}\, \rho^3 
\left[  {1 \over \rho^2 + |X|^2} |D X|^2  
 +  {\Delta m^2 \over \rho^2} |X|^2   + {1 \over 2} F_V^2  \right], 
\label{daq}
\end{equation}
The  field $X$ (the equivalent of the embedding coordinates $L,\phi$ of the D7 brane in the top down model) describes
the quark condensate degree of freedom. Fluctuations in $|X|$ around its vacuum configurations describe the scalar meson. The $\pi$ fields are the phase of $X$,
\begin{equation} X = L(\rho)  ~ e^{2 i \pi^a T^a} .
\end{equation}
$F_V$ are vector fields that will describe the vector ($V$)  mesons.

We work with the five dimensional metric 
\begin{equation} 
ds^2 =  { d \rho^2 \over (\rho^2 + |X|^2)} +  (\rho^2 + |X|^2) dx^2, 
\end{equation}
which will be used for contractions of the space-time indices.
$\rho$ is the holographic coordinate 
and $|X|=L$ enters into the effective radial coordinate in the space, i.e. there is an effective $r^2 = \rho^2 + |X|^2$. This is how the quark condensate generates a soft IR wall for the linearized fluctuations that describe the mesonic states: when $L$ is non-zero the theory will exclude the deep IR at $r=0$.


The normalizations of $X$ and $F_V$ are determined by matching to the gauge theory in the UV. External currents are associated with the non-normalizable modes of the fields in AdS. In the UV we expect 
$|X| \sim 0$ and we can solve
the equations of motion for the scalar, $L= K_S(\rho) e^{i q.x}$ and vector $V^\mu= \epsilon^\mu K_V(\rho) e^{i q.x}$     field. Each satisfies the same equation
\begin{equation}  \label{thing}
\partial_\rho [ \rho^2 \partial_\rho K] - {q^2 \over \rho} K= 0\,. \end{equation}
The UV solution  is
\begin{equation} \label{Ks}
K_i = N_i \left( 1 + {q^2 \over 4 \rho^2} \ln (q^2/ \rho^2) \right),\quad (i=S,V),
\end{equation}
where $N_i$ are normalization constants that are not fixed by the linearized equation of motion.
Substituting these solutions back into the action gives the scalar correlator $\Pi_{SS}$ and the vector correlator $\Pi_{VV}$. Performing the usual matching to the UV gauge theory requires us to set
\begin{equation} N_S^2 = N_V^2 = {N_c N_f \over 24 \pi^2 }.
\end{equation}

The vacuum structure of the theory can be determined by setting all fields except $|X|=L$ to zero. We assume that $L$ will have no dependence on the $x$ coordinates. The action for $L$  is given by
\begin{equation} \label{act} S  =  \int d^4x~ d \rho ~  \rho^3 \left[   (\partial_\rho  L)^2 +  \Delta m^2 {L^2  \over \rho^2 }   \right].
\end{equation}
If $\Delta m^2 =0$ then the scalar, $L$, describes a dimension 3 operator and dimension 1 source as is required for it to represent $\bar{q} q$ and the quark mass $m$. That is, in the UV the solution for the $L$ equation of motion is $L = m + \bar{q}q/\rho^2$. A non-zero $\Delta m^2$ allows us to introduce an anomalous dimension for this operator. If the mass squared of the scalar violates the BF bound of -4 ($\Delta m^2=-1$, $\gamma=1$) then  the scalar field $L$ becomes unstable and the theory enters a chiral symmetry breaking phase.

We will fix the form of $\Delta m^2$ using the two loop running of the gauge coupling in QCD with $N_f$ flavours transforming under a representation $R$. This takes the form
\begin{equation} 
\mu { d \alpha \over d \mu} = - b_0 \alpha^2 - b_1 \alpha^3,
\end{equation}
where
\begin{equation} b_0 = {1 \over 6 \pi} \left(11 C_2(G) - 4N_fC_2(R)\frac{\text{dim}(R)}{\text{dim}(G)}\right), \end{equation}
and
\begin{equation}b_1 = {1 \over 8 \pi^2} \left(\frac{34}{3}\left[C_2(G)\right]^2-\left[\frac{20}{3}C_2(G)C_2(R)+4\left[C_2(R)\right]^2\right]N_f\frac{\text{dim}(R)}{\text{dim}(G)}\right) \, . \end{equation}

Above, we denote the adjoint representation as $G$ and its respective Casimir by $C_2(G)=N_c$. Table \ref{table:casreps} shows all the distinguishing quantities associated to each of the representations we consider: the dimension of the representation, $C_2(R)$, and the minimum number of flavours required for loss of asymptotic freedom, $N_f^\text{max}$.
\begin{table}[ht]
\centering
\begin{tabular}{|c|c|c|c|}\hline
 $R$&dim($R$)&$C_2(R)$&$N_f^\text{max}$\\
 \hline &&&\\
 Fundamental & $N_c$ & $\frac{N_c^2-1}{2N_c}$  & $\frac{11}{2}N_c$\\&&&\\
 Adjoint (G) & $N_c^2-1$&$N_c$ &$2\frac{3}{4}$\\&&&\\
 2IS &$\frac{N_c(N_c+1)}{2}$ &$\frac{(N_c-1)(N_c+2)}{N_c}$ &$\frac{11}{2}\frac{N_c}{N_c+2}$\\&&&\\
 2IA &$\frac{N_c(N_c-1)}{2}$ &$\frac{(N_c+1)(N_c-2)}{N_c}$ &$\frac{11}{2}\frac{N_c}{N_c-2}$\\&&&\\
 \hline
\end{tabular}
\caption{Distinguishing quantities of representations of $SU(N_c)$ gauge theories with asymptotic 
freedom valid for any $N_c\geq2$.}
\label{table:casreps}
\end{table}

The one loop result for the anomalous dimension of the quark mass is
\begin{equation} \gamma_1(\mu;R) = {3 C_2(R) \over 2\pi}\alpha(\mu;R).  \end{equation}

We will identify the RG scale $\mu$ with the AdS radial parameter $r = \sqrt{\rho^2+L^2}$ in our model. Note it is important that $L$ enters here. If it did not and the scalar mass was only a function of $\rho$ then, were the mass to violate the BF bound at some $\rho$, it would leave the theory unstable however large $L$ grew. Including $L$ means that the creation of a non-zero but finite $L$ can remove the BF bound violation leading to a stable solution. This is analogous to what happens in the top-down model.

Working perturbatively from the AdS result $m^2 = \Delta(\Delta-4)$ we have
\begin{equation} \label{dmsq3} \Delta m^2 = - 2 \gamma_1(\mu;R) = -{3 C_2(R) \over \pi}\alpha(\mu;R).\end{equation}
This will then fix the $r$ dependence of the scalar mass through $\Delta m^2$ as a function of $N_c$ and 
$N_f$ for each $R$.  Note that if one were to attempt such a matching beyond two loop order the perturbative result would become gauge dependent. We hope that the lower order gauge independent results provide sensible insight into the running in the theory.

It is important to stress that using the perturbative result outside the perturbative regime is in no sense rigorous, but simply a phenomenological parametrization of the running as a function of $\mu, N_c,N_f$ that shows fixed-point behaviour. We expect broad trends in the behaviour of the theories with varying $N_f, N_c$ to be sensibly described with this ansatz. Similarly, the relation \eqref{dmsq3} between $\Delta m^2$ and $\gamma_1$ is a guess outside of the perturbative regime. Note that the holographic fixed point value for the anomalous dimension is given by solving $\Delta(\Delta-4)= \Delta m^2$ and the resultant $\gamma$ will not be the same as the fixed point in $\gamma_1$ away from the perturbative regime.

The vacuum structure for a given choice of representation, $N_f$ and $N_c$ must be identified first. The Euler-Lagrange equation for the  vacuum embedding $L_v$ is given at fixed $\Delta m^2$ by the solution of 
\begin{equation}\label{embedeqn}
 \frac{\partial}{\partial\rho}\left( \rho^3 \partial_\rho L_v\right)  - \rho \Delta m^2 L_v =0.
\end{equation}
Note that if $\Delta m^2$ depends on $L$ at the level of the Lagrangian then there would be an additional term  $- \rho L_v^2 \partial \Delta m^2 / \partial L_v$. We neglect this term and instead impose the running of $\Delta m^2$ at the level of the equation of motion. The reason is that the extra term introduces an effective contribution to the running of $\gamma$ that depends on the gradient of the running coupling. Such a term is not present in perturbation theory in our QCD-like theories - we wish to keep the running of $\gamma$ in the holographic theory as close to the perturbative guidance from the gauge theory as possible.

In order to find $L_v(\rho)$ we solve the equation of motion numerically with shooting techniques with an input IR initial condition. A sensible first guess for the IR  boundary condition is
\begin{equation}\label{bca}  L_v(\rho=L_0) = L_0, \hspace{1cm}  L_v'(\rho=L_0)=0. \end{equation}
This IR condition is similar to that from top down models\cite{Karch:2002sh} but imposed at the RG scale where the flow becomes ``on-mass-shell". Here we are treating $L_v(\rho)$ as a constituent quark mass at each scale $\rho$. Were we to continue the flow below this quark mass scale we would need to address the complicated issue of the decoupling of the quarks from the running function $\gamma$. 
\bigskip

\subsection{Meson Spectra}
We now turn to computing the physical parameters, the masses of the $(\rho,\sigma,\pi)$-mesons and the scalar glueball, for each viable representation. These parameters are true predictions of the model which, just as in the gauge theories,  depend only on the choice of the quark mass, $N_c, N_f$ and the scale $\Lambda$. 
 \subsection{Linearized Fluctuations} 

The isoscalar $\bar{q}q$ ($\sigma$) mesons are described by linearized fluctuations of $L$ about its vacuum configuration, $L_v$. We look
for space-time dependent excitations,  ie $|X| = L_v + \delta(\rho) e^{i q.x}$,  $q^2=-M_\sigma^2$. The equation of motion for $\delta$ is, linearizing (\ref{embedeqn}),
\begin{equation} \label{deleom} \partial_\rho( \rho^3 \delta' ) - \Delta m^2 \rho \delta -   \rho L_v \delta \left. \frac{\partial \Delta m^2}{\partial L} \right|_{L_v} 
+ M_\sigma^2 R^4 \frac{\rho^3}{(L_v^2 + \rho^2)^2} \delta  = 0\,.  \end{equation}
We seek solutions with, in the UV, asymptotics of $\delta=\rho^{-2}$ and with $\partial_\rho\delta|_{L_0}=0$ in the IR, giving a discrete meson spectrum.

The isovector ($\rho$) meson spectrum is determined from the normalizable solution of the equation of motion for the spatial pieces of the vector gauge field $V_{\mu \perp} = \epsilon^\mu V(\rho) e^{i q.x}$ with $q^2=-M^2$. The appropriate equation is
\begin{equation} \label{vv}  \partial_\rho \left[ \rho^3 \partial_\rho V \right] + {\rho^3 M^2 \over (L_v^2 + \rho^2)^2} V = 0\,. \end{equation}
We again impose $\partial_\rho V|_{L_0}=0$ in the IR and require in the UV that $V\sim c/\rho^2$. 

The pion mass spectrum is identified by assuming a space-time dependent phase $\pi^a(x)$ of the AdS-scalar $X$ describing the $\bar{q}q$ degree of freedom, i.e $X=L(\rho)\exp(2i\pi^a(x)T^a)$. The equation of motion of the pion field is then,
\begin{equation}
\partial_\rho\left(\rho^3L_v^2\partial_\rho\pi^a\right)+M_\pi^2\frac{\rho^3L_v^2}{(\rho^2+L_v^2)^2}\pi^a=0.
\label{pionfield}
\end{equation}
Again, we impose at the IR boundary that $\partial_\rho\pi^a|_{L_0}=0$.

\subsection{Results}
We can now move to displaying the outcomes of the Dynamic AdS/QCD theory. We will fix the strong coupling scale $\Lambda$ by fixing the $\rho$ mass at $M_\pi=0$ for each choice of representation, $N_f$ and $N_c$ and express all quantities in units of that scale. For our  plots then the only input parameters are the quark mass, $N_f$ and $N_c$. We will explore a range of gauge theories with different quark matter. 

\begin{figure}[]
\centering
\includegraphics[width=16cm]{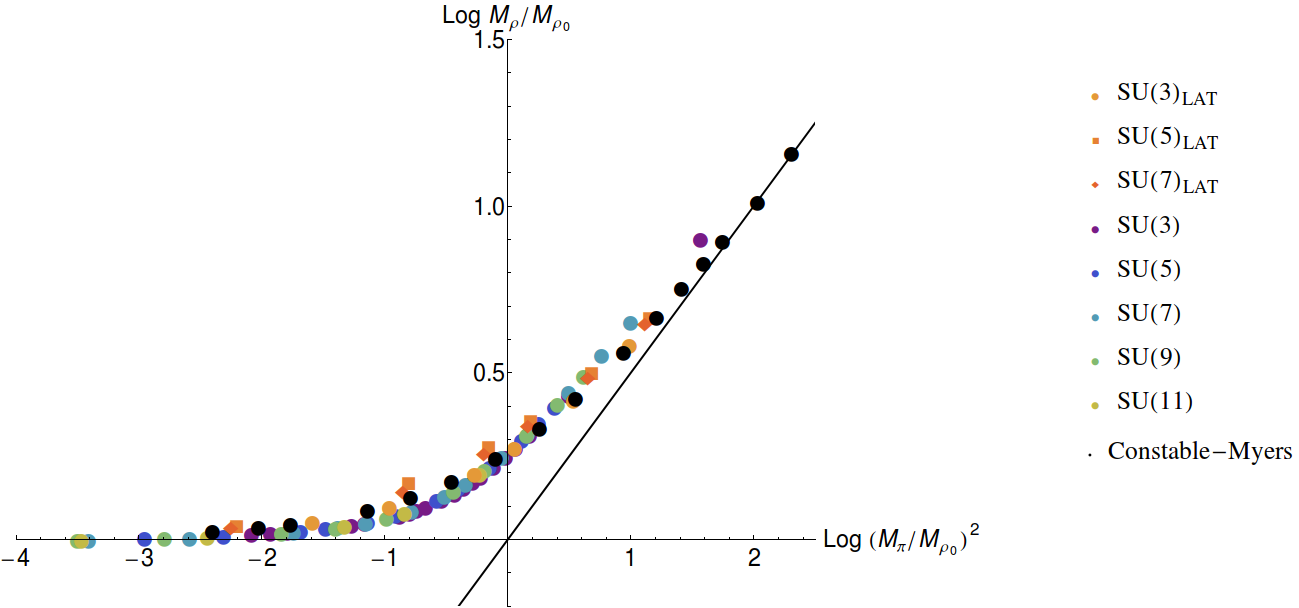}
\caption{A Log-Log plot of $M_\rho$ versus $M_\pi^2$ - the plot displays the quenched lattice data from \cite{Bali:2013kia}, the top-down Constable-Myers model of section 2 and the quenched results for varying $N_c$ in Dynamic AdS/QCD from section 3. The solid line corresponds to $M_\rho=M_\pi$.}
\label{nf=3fund}
\end{figure}

\subsubsection{Quenched Fundamental Representation}

To test the model, we first  compute  $M_\rho$ and $M_\pi$ in the model with quenched fundamental quarks. This means that we do not  include the quark contribution in the running of the gauge coupling. We compute the meson masses as functions of $N_c$ to compare with the previously discussed quenched lattice data of Fig \ref{CMvslat}. We display the data also in  Fig \ref{CMvslat}. We note that all choices of SU($N_c$) give essentially the same curve in this plot. This curve lies below, but within $5\%$ of the prediction of the Constable-Myers top-down model.  The result for the Dynamic AdS/QCD model in this plot  displays some curvature over the range of the lattice data, suggesting that the linear extapolation used to place the lattice data on the plot maybe incorrect. This suggests the results for the lattice data in our linear fit are slightly too large, by as much as $5 \%$.  
{Indeed in \cite{Bali:2013kia} evidence is presented for a non-linear fit already in the lattice data.}
Given the expectation of some systematic error on the lattice data (see \cite{Bali:2013kia}) the match between all these models is remarkable and lends considerable support to further predictions of the Dynamic AdS/QCD model.

To emphasise how well the results match,  we also plot the same Dynamic AdS/QCD and lattice data on a Log-Log plot in Fig \ref{nf=3fund}. The figure also displays the line $M_\rho=M_\pi$, which would be the one appropriate to a very weakly coupled theory where both mesons masses are just twice the quark mass. This line is expected  to be approached  at large $M_\pi$, i.e.~in the limit of large quark mass.  Clearly, the very different computations for these theories agree rather well. Whilst both the holographic model's curves are compatible with the lattice data at the level of the errors due to the coarse lattice spacing taken in \cite{Bali:2013kia}, the top-down Constable Myers model does fit the data mildly better (the $M_\rho$ points are raised by upto $2\%$ or so), including in the large $M_\pi$ limit. If this is indeed the case, then it is likely due to  $\gamma$ in that model falling to zero more quickly than in QCD as function of the RG scale  - the holographic description of the UV is probably closer to perturbative QCD with $\gamma=0$.

\subsubsection{Fundamental Representation} 

\begin{figure}[]
\centering
\includegraphics[width=16cm]{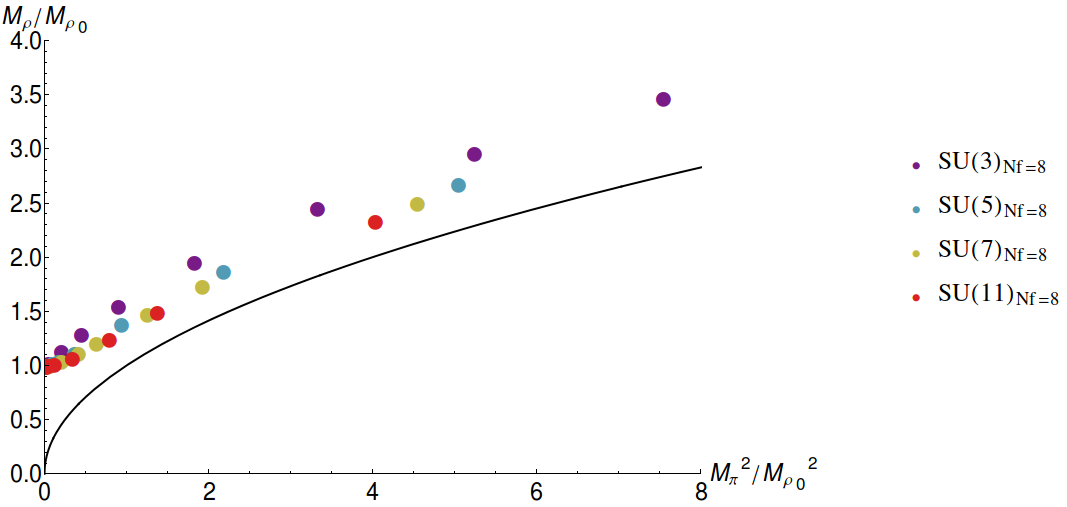}\bigskip\bigskip\bigskip\bigskip\bigskip

\includegraphics[width=16cm]{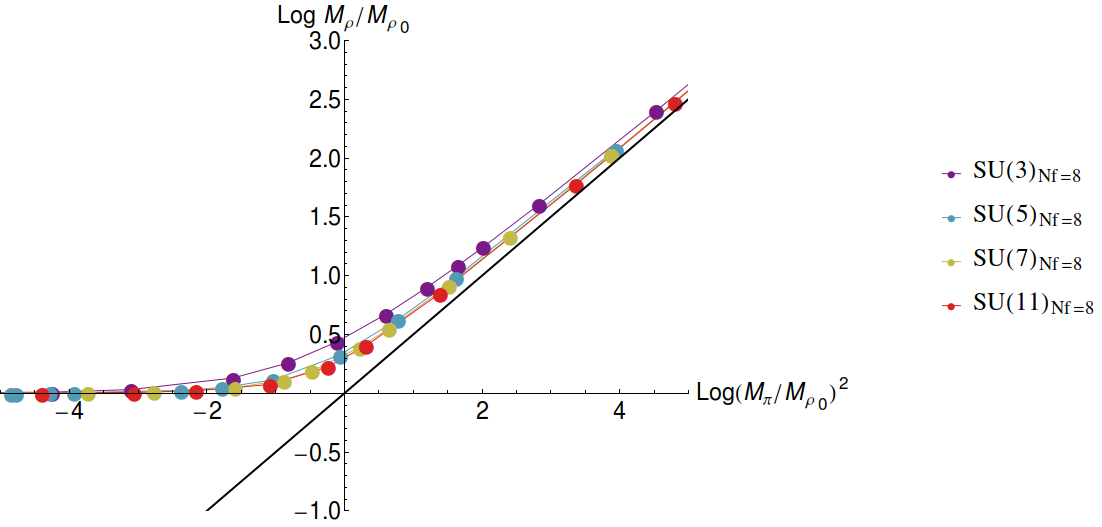}
\caption{$M_\rho$ versus $M_\pi^2$ in SU(N$_c$) theory with $N_F=8$ fundamental quarks - the lower plot shows the same in Log Log format. The solid line corresponds to $M_\rho=M_\pi$.}
\label{nf=8fund}
\end{figure}

The quenched results display very little dependence on $N_c$. The reason is that the running of $\gamma$ at the point $\gamma=1$ is very fast in all these cases so the dynamics comes out very similar. To see some $N_c$ dependence we should unquench the theory and include a sufficent number of quarks to affect the running. For example in Fig \ref{nf=8fund}
we show the $N_c$ variation in the $M_\rho$ - $M_\pi^2$ plane of a theory with $N_f=8$.  The dependence on $N_c$  is again not huge but for low $N_c$ there is a clear distinction from larger $N_c$ theories that are effectively more quenched. This further emphasises the success of the holographic model in lying so close to the quenched lattice data - it is not that the curves seen in the previous subsection are the only outcome!

\begin{figure}[]
\centering
\includegraphics[width=16cm]{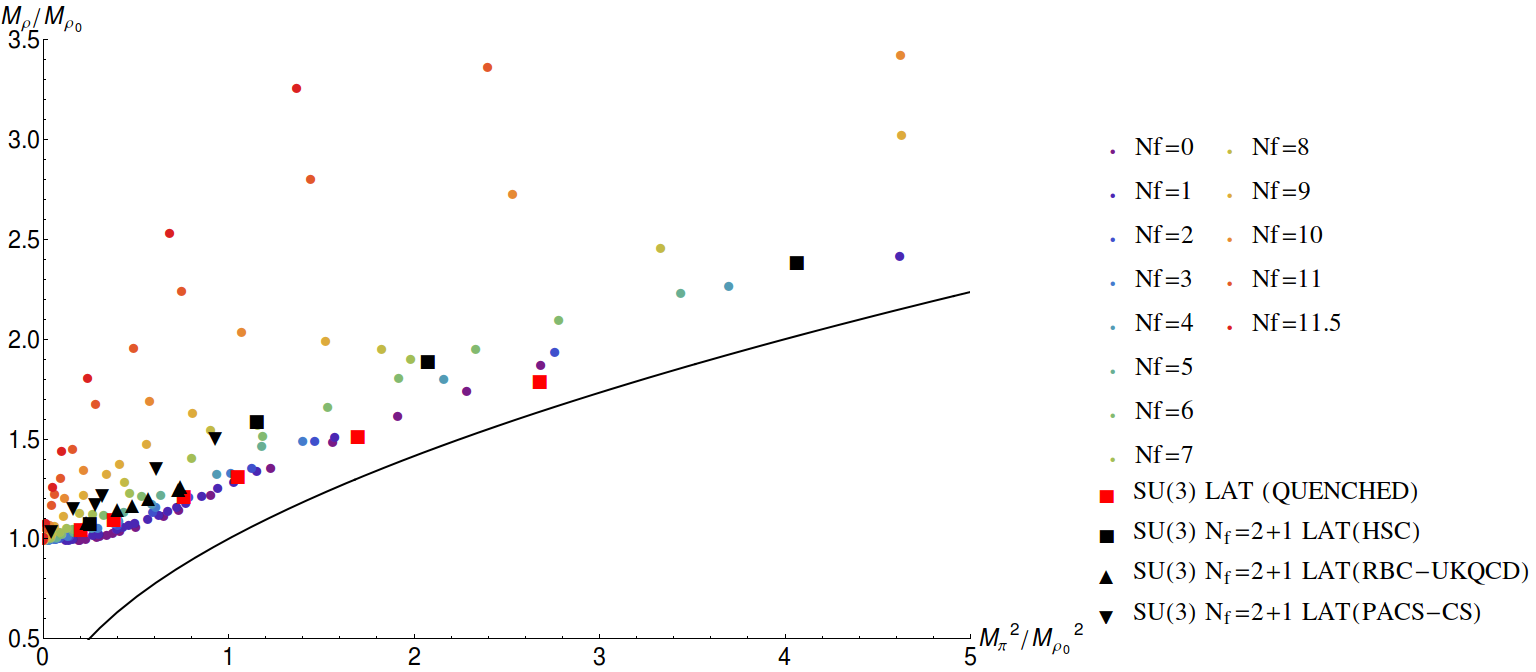}\bigskip\bigskip\bigskip\bigskip\bigskip

\includegraphics[width=16cm]{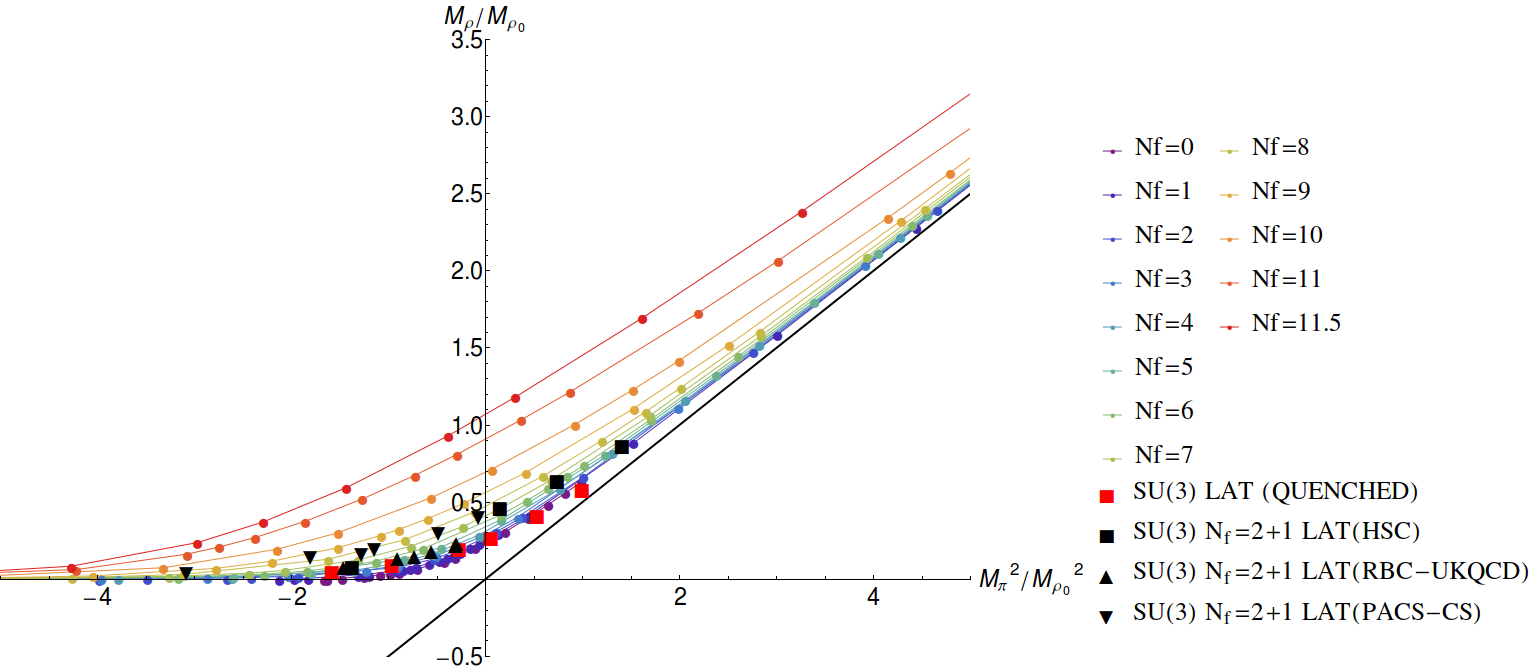}
\caption{ SU(3) gauge theory with $N_f$ fundamental quarks showing the approach to the conformal window at $N_f=12$. The lower plot is a Log Log version of the top plot. The solid line corresponds to $M_\rho=M_\pi$. The plots also show
lattice data for the quenched theory \cite{Bali:2013kia} and unquenched $N_f=3$ theory \cite{Lin:2008pr,Aoki:2008sm,Allton:2007hx}. }
\label{su3fund}
\end{figure}

We can now turn to study the question of whether there are choices of $N_f$ and $N_c$  that provide spectra very different from QCD-like theories. As is well known, the theories that are most unlike QCD are those on the edge of the conformal window. For example, for an SU(3) gauge theory at $N_f > 12$ the running of $\gamma$ flows to an IR fixed point below $\gamma=1$. In the holographic model the BF bound is not violated and chiral symmetry breaking does not occur. For theories that do break chiral symmetry at slightly lower values of $N_f$ the theory runs close to an IR fixed point that just violates the BF bound. Crucially at the point where $\gamma=1$ the gradient of the running of $\gamma$ is much smaller than in QCD-like theories. To demonstrate the impact of this on the spectrum we plot the $N_f$ dependence of the SU(3) theory in the $M_\rho$ - $M_\pi^2$ plane in Fig. \ref{su3fund}. The $\rho$ mass is substantially enhanced relative to the $\pi$ mass at larger $N_f$.

Theories with slow running at the scale where chiral symmetry breaking is triggered are called walking gauge theories
\cite{Holdom:1981rm}. For these theories the chiral condensate has dimension approximately 2 at the IR scale of conformal symmetry breaking. The dimension 3 UV condensate is roughly the product of that dimension 2 IR condensate and the scale of the 1-loop $\beta$ function of the theory that determines where the anomalous dimension changes from $\gamma=0$ to $\gamma=1$. This scale will be much larger than the IR scale and the UV condensate enhanced. The usual expectation is that the $\rho$ mass will be proportional to $\langle \bar{q} q \rangle^{1/3}$ whilst the $\pi$ mass will scale as $m_q^{1/2} \langle \bar{q} q \rangle^{1/6}$. An enhancement of the condensate would therefore raise $M_\rho$ at any fixed $M_\pi$ as is seen in  Fig. \ref{su3fund}.  Generically for different $N_c$ we observe the same behaviour as $N_f/N_c \rightarrow 4$.

This is a good point to compare our Dynamic AdS/QCD theory to unquenched lattice data \cite{Lin:2008pr,Aoki:2008sm,Allton:2007hx}. We have seen that the effect of including more quarks in our model is that the value of $M_\rho$ rises at fixed $M_\pi$. This suggests that the effect of quark loops is to raise $M_\rho$. We display lattice data in the top plot of Fig. \ref{su3fund} - we show both the quenched results previously discussed for SU(3) gauge theory, but now also unquenched data for the same theory with $N_f=3$, taken from \cite{Lin:2008pr,Aoki:2008sm,Allton:2007hx}. The three sets of lattice data show some spread in the low $M_\pi$ region, but we indeed observe a shift upwards in $M_\rho$ by 20$\%$ or so. In fact, the fit to the Dynamic AdS/QCD model for $N_f=3$ is a little poorer than to the quenched lattice data - the lattice points are more similar to the $N_f=5$ version of Dynamic AdS/QCD (although there is clearly some uncertainty in the lattice results as shown by the spread). This is most plausibly explained as a failure of the very naive perturbative based running ansatz we have used as an input into the model. The key measure is the gradient of $\gamma$ with RG scale at the scale where $\gamma=1$. For $N_f=3$ $\gamma'=-4.25$, whilst at $N_f=5$ $\gamma' = -3.70$. This implies that the shift in spectrum is  caused by a 15$\%$ shift in this gradient. Clearly the perturbative ansatz cannot be trusted at this level of accuracy. It is not surprising that the precise features of the spectrum are dependent on the choice of assumed running for $\gamma$. It is encouraging that the holographic model correctly gets gross features correct, such as the rise in $M_\rho$ in theories with more quark loops. This gives us confidence that the holographic model can be useful in understanding broad trends in the spectrum as quark content of the theory is changed.

\begin{figure}[]
\centering
\includegraphics[width=16cm]{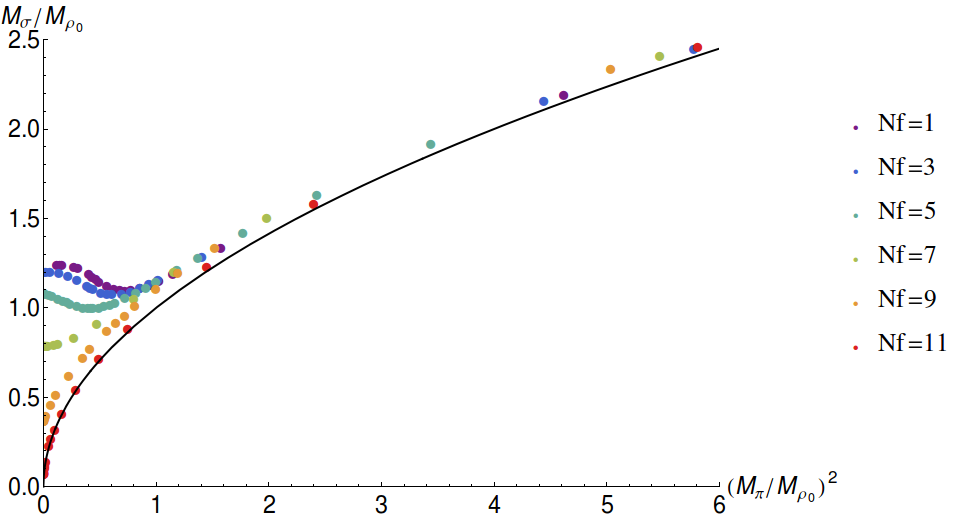}\bigskip\bigskip\bigskip\bigskip\bigskip

\includegraphics[width=16cm]{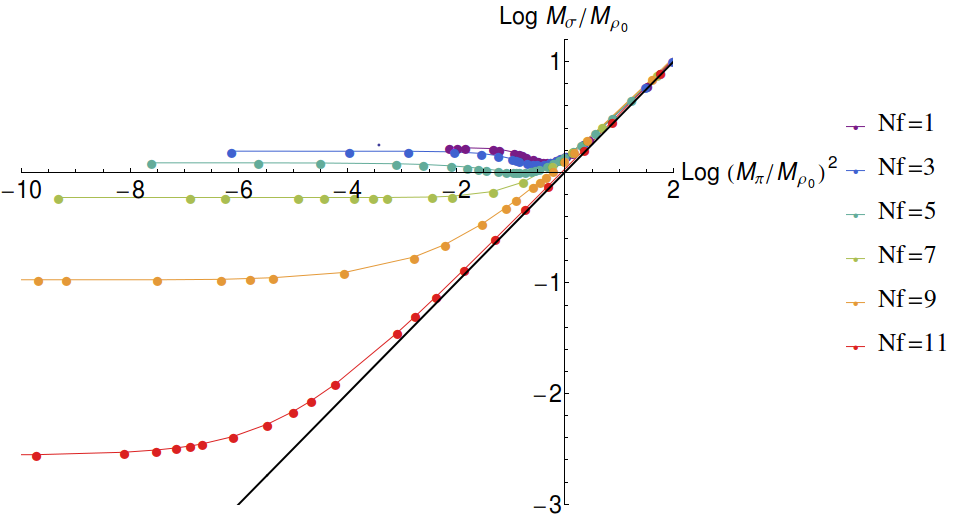}
\caption{$M_\sigma$ versus $M_\pi^2$ in SU(3) gauge theory with varying $N_f$ fundamental quarks. The lower plot is a Log Log version of the top plot. The solid line corresponds to $M_\sigma=M_\pi$.}
\label{sigmasu3fund}
\end{figure}

An additional expectation in a walking theory is that the $\sigma$ mode $\bar{q} q$ bound state should become light as one approaches the edge of the conformal window from below. The reason is that since the quark condensate is enhanced the effective potential for the condensate becomes flatter as discussed in \cite{Haba:2010hu}. To observe this let us now turn to computing the $\sigma$ meson mass. We will again pick $N_c=3$ as an example and show the $N_f$ dependence of $M_\sigma$ against $M_\pi^2$ in Fig. \ref{sigmasu3fund}. The $N_f=7$ curve is perhaps what one would have predicted for QCD - at large quark mass the $\sigma$ and $\pi$ masses become degenerate. At low quark mass as the $\pi$ mass tends to zero the $\sigma$ mass saturates at a value below the $\rho$ mass. One might then identify this state with the $f_0(500)$ state observed in experiment.  However, for $N_f=3$ the holographic model predicts that the lightest $\sigma$ is heavier than the $\rho$ and it looks more sensible to match it to the $f_0(980)$ which it matches at the 10$\%$ level. An explanation of the origin of the lighter $f_0$ would then be needed. In fact though the literature has considerable speculation about this state which might be a molecule or some other exotic state (see for example \cite{Londergan:2013dza}). We can not resolve this issue here. However, the main use of our model is to look at significant trends in the behaviour of the spectra as we adjust the running of $\gamma$. Here our plot very strongly supports the speculation that this $\sigma$ mode becomes light as one approaches the walking regime and the edge of the conformal window at $N_f=12$.

\newpage

\subsubsection{Other Representations}

As we have stressed above Dynamic AdS/QCD can accommodate a description of any arbitrary quark representation. The flavour representation enters through the running of the anomalous dimension $\gamma$ (for which we continue to use the two loop perturbative result). In this section we provide some plots showing some exploration of the larger space of theories.  

\begin{figure}[]
\centering
\includegraphics[width=13cm]{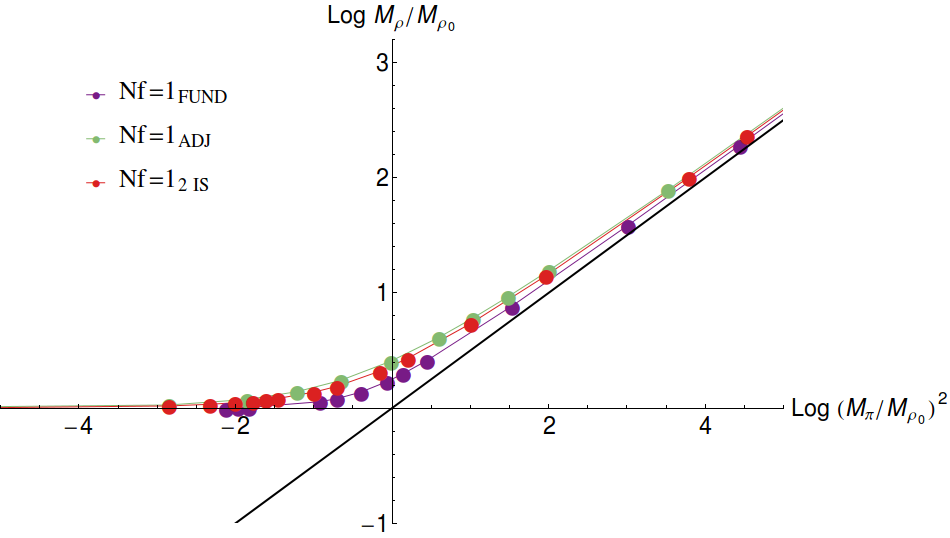} \bigskip

\includegraphics[width=13cm]{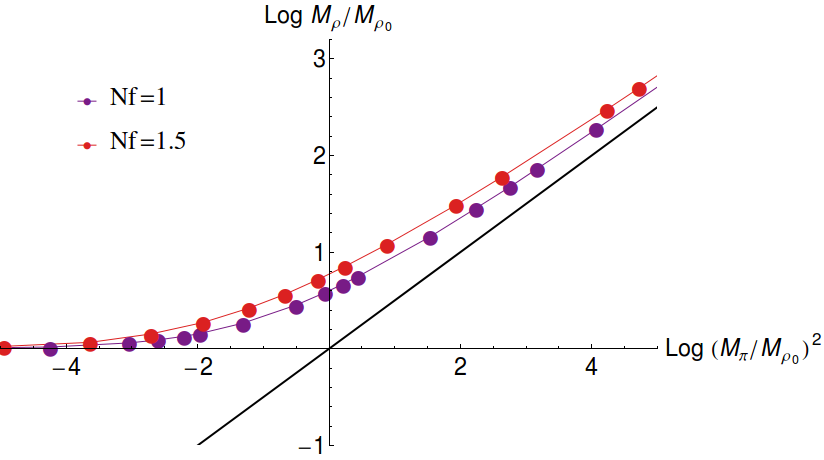} \bigskip

 \includegraphics[width=13cm]{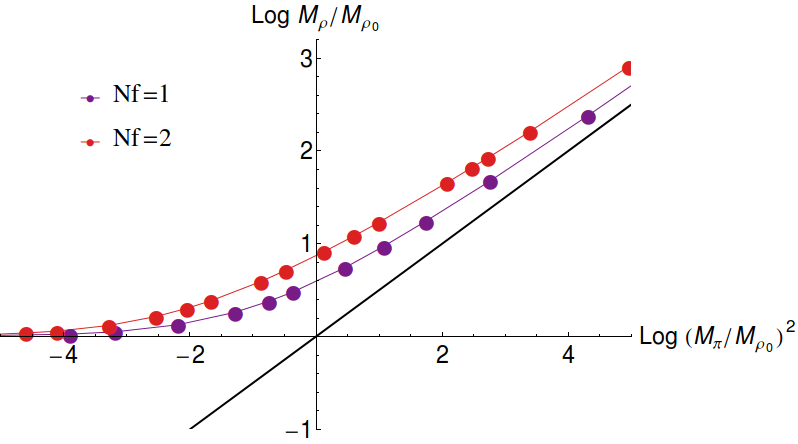}
\caption{A Log-Log plot in the $M_\rho$-$M_\pi^2$ plane for SU(3) gauge theory. The top plot shows the results in models with $N_f=1$ but with the fermions in the fundamental, adjoint and 2-index symmetric representations. The middle figure shows the $N_f$ dependence in the case with adjoint fermions and the bottom plot the same for the 2-index symmetric representation. The solid line corresponds to $M_\sigma=M_\pi$.} 
\label{su3}
\end{figure}

As a first example in Fig. \ref{su3} we show results for $N_c=3$. The top plot shows the results in the $M_\rho$ vs $M_\pi^2$ plane for the theory with a single quark in the fundamental representation (here the same as the 2 index anti-symmetric representation), the adjoint representation, and the 2 index symmetric representation. Increasing the size of the representation makes a bigger impact on the running of the coupling and moves the curve away from QCD-like towards the walking regime. In the lower two plots we show the $N_f$ dependence for the adjoint and 2 index symmetric representation (here we allow $N_f=1.5$ since by $N_f=2$ chiral symmetry breaking is lost). Adding flavours makes the theory more walking in behaviour. 

\begin{figure}[]
\centering
\includegraphics[width=16cm]{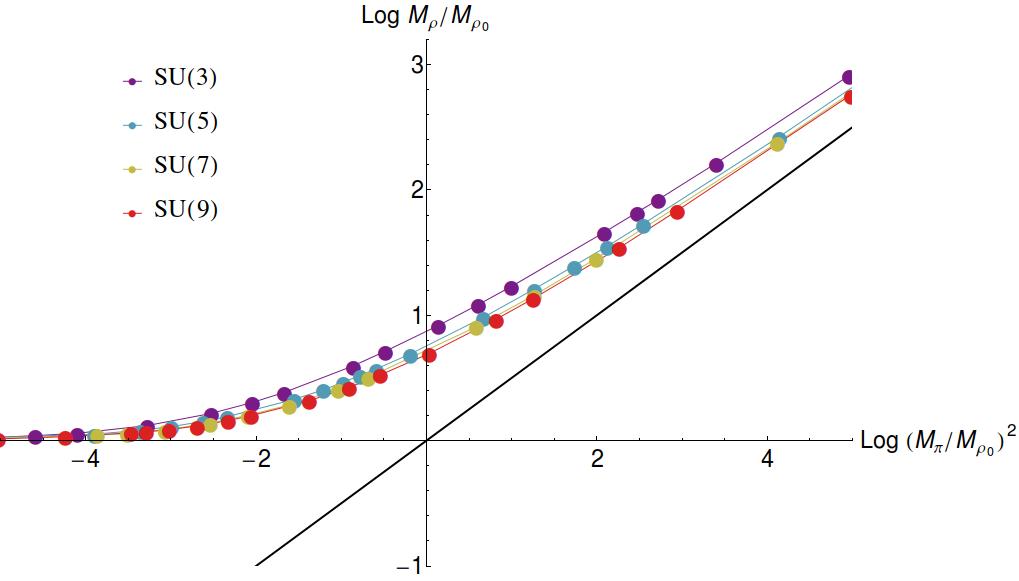}
\caption{A Log-Log plot in the $M_\rho$-$M_\pi^2$ plane for SU($N_c$) gauge theory with $N_f=2$ 2-index symmetric representation quarks.  The solid line corresponds to $M_\sigma=M_\pi$.}
\label{nf=22is}
\end{figure}

We can also explore the $N_c$ dependence of these theories at fixed $N_f$. For example in Fig. \ref{nf=22is} we vary $N_c$ with two 2 index symmetric representation quarks. Increasing $N_c$ moves the theory closer to the quenched limit and a more QCD-like spectrum. Within this space of theories we are not finding any additional structure beyond the dependence on the rate of running at the point $\gamma=1$.

\begin{figure}[]
\centering
\includegraphics[width=16cm]{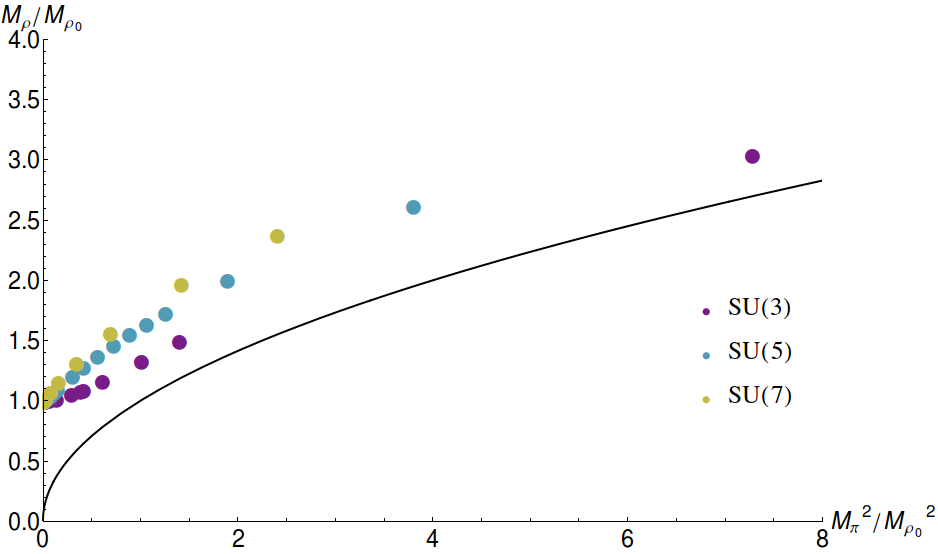}
\caption{A Log-Log plot in the $M_\rho$-$M_\pi^2$ plane for SU($N_c$) gauge theory with $N_f=3$ 2-index anti-symmetric representation quarks.  The solid line corresponds to $M_\sigma=M_\pi$.}
\label{nf=32ias}
\end{figure}

One final interesting case is that of 2 index anti-symmetric representation quarks. As one moves to higher $N_c$ at fixed $N_f$ the two loop IR fixed point value of the coupling actually decreases. For these theories increasing $N_c$ moves one towards the walking regime. We show this in Fig. \ref{nf=32ias}.

\begin{figure}[]
\centering
\includegraphics[width=13cm]{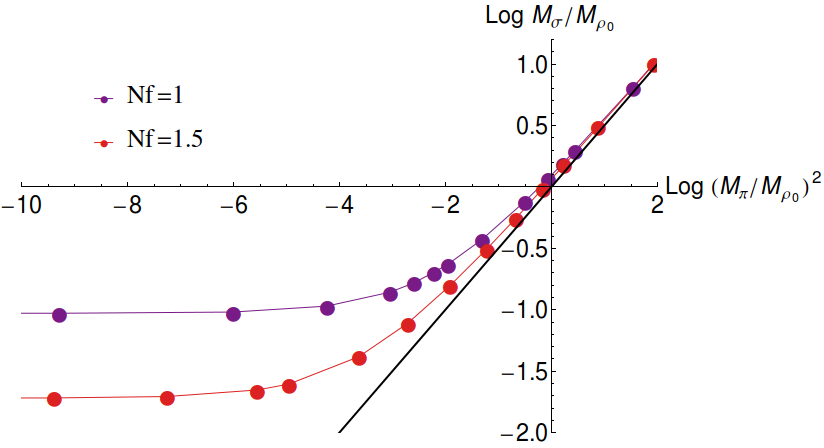}\bigskip

\includegraphics[width=13cm]{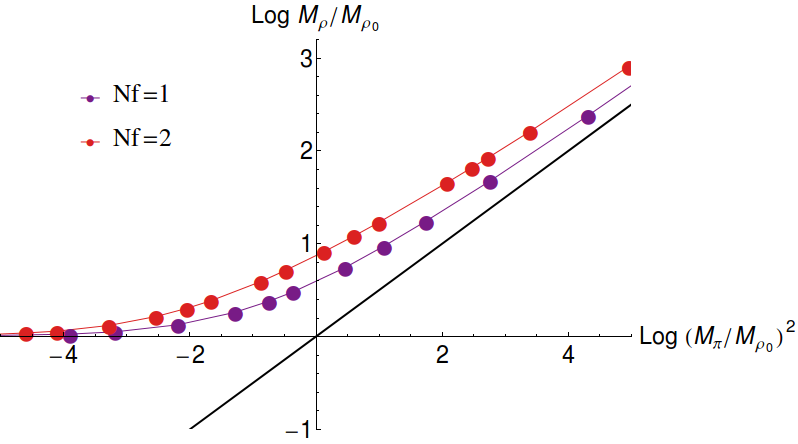} \bigskip

\includegraphics[width=13cm]{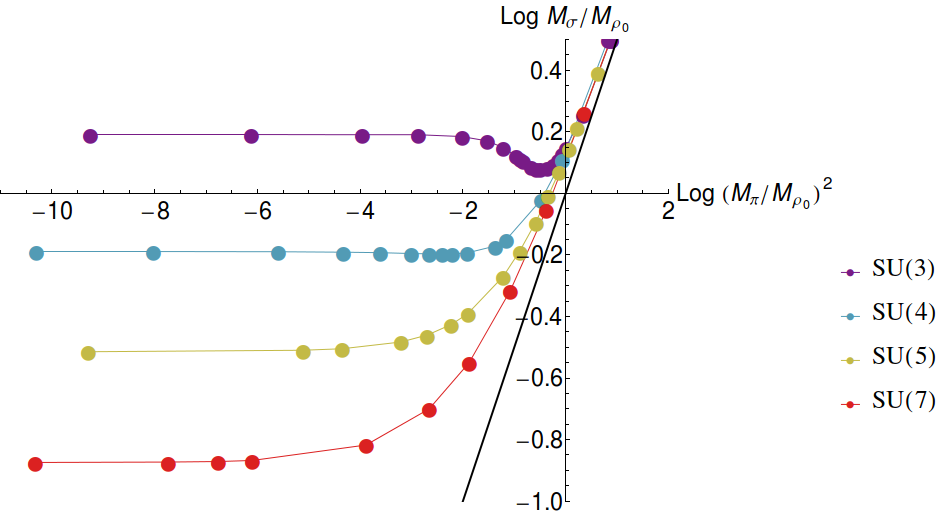}
\caption{Log-Log plots in the $M_\sigma$-$M_\pi^2$ plane. The top plot shows the results in SU(3) gauge theory with adjoint quarks. The middle plot is for SU($N_c$) gauge theory with $N_f=2$ 2-index symmetric representation quarks. The bottom plot is for SU($N_c$) gauge theory with $N_f=3$ 2-index anti-symmetric representation quarks. The solid line corresponds to $M_\sigma=M_\pi$.}
\label{sigmavar}
\end{figure}

The walking regimes of these theories also display a light $\sigma$ meson. We show this trend for a variety of sequences of theories moving towards the walking regime in Fig. \ref{sigmavar}. The trends in the spectrum as one approaches the walking regime across a wide range of theories are very similar.

$\left. \right.$

\newpage

\subsection{The Scalar Glueball}


Another state that one might be interested in studying as part of the lightest spectra of these theories is the lightest glueball state (see \cite{DelDebbio:2010ze} for some discussions in preliminary lattice simulations). AdS/QCD is not suited to study this state since it is fundamentally a description of the quark sector. The glueball could be included as a separate scalar in AdS but one would then need to correctly encode its dynamics for the gauge theory's vacuum Tr$F^2$ condensate and make a guess as to how it couples in the scalar potential to the quark condensate field $X$. There are  a lot of unknown parameters that describe the mixing of the $\sigma$ and glueball state. Rather than attempt this here we will instead make a back of the envelope computation for the glueball state. 

In pure Yang Mills the glueball is expected to be between 5 and 10 times the one loop strong coupling scale. In the Dynamic AdS/QCD model we have asumed the two loop running for the gauge coupling and $\gamma$ and then computed the IR quark mass gap, the value of $L$ at the on-mass shell condition. A simple thing to do then is to decouple the quarks at that scale $L_{\rm on-mass}$ and use the one loop pure Yang Mills coupling into the IR. We compute the position of the IR pole and multiply by 8 to estimate the glueball mass. This will at least give a ball-park behaviour although mixing is explicitly not addressed. 

\begin{figure}[]
\centering
\includegraphics[width=16cm]{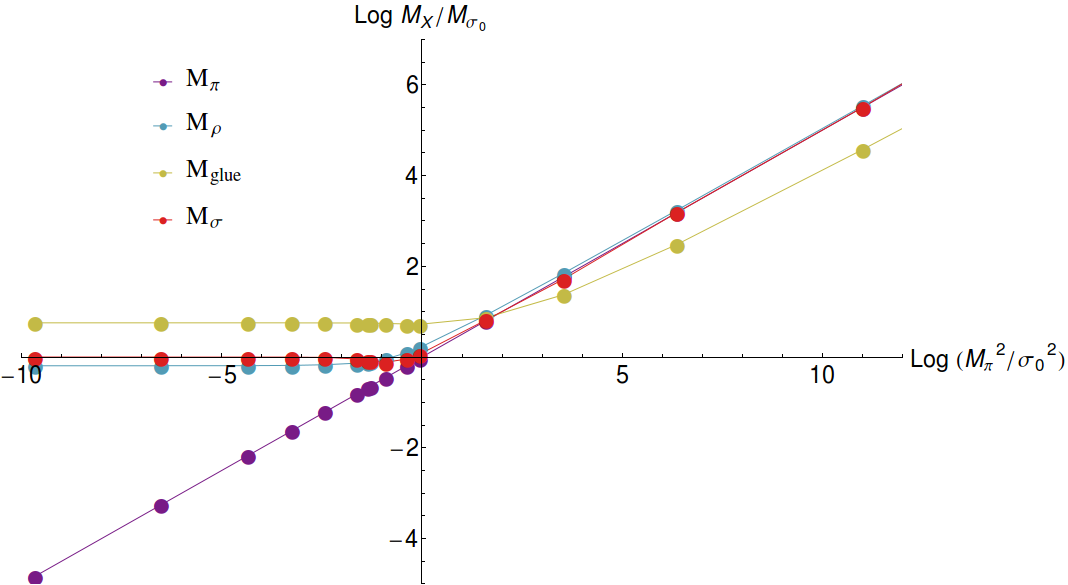}\bigskip\bigskip\bigskip\bigskip\bigskip

\includegraphics[width=16cm]{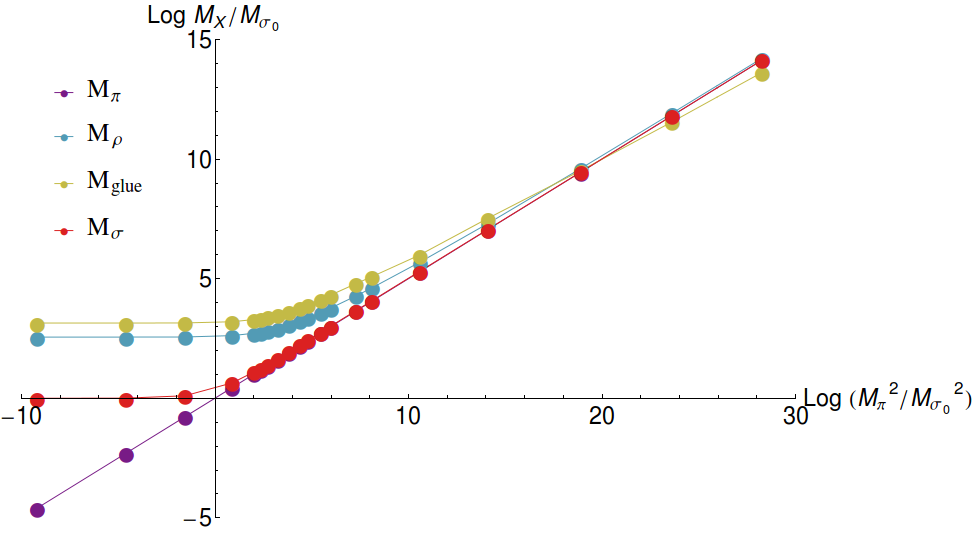}
\caption{The spectra of the $N_c=3$ gauge theory with fundamental quarks - the top figure shows $N_f=3$, the bottom  $N_f=11$.}
\label{glu}
\end{figure}

In Fig. \ref{glu} we display the spectra of the $N_c=3$ theory for $N_f=3$ (QCD-like) and $N_f=11$ (close to walking) including the glueball. Both theories display a Goldstone  pion. As we have seen before the $\sigma$ becomes light and interchanges ordering with the $\rho$ as one approached the walking regime.  In both cases the glueball is the lightest state at large quark mass - here the quarks decouple at their mass scale, where the glue is still weakly coupled, and the pure glue theory then runs logarithmically to strong coupling at a much lower scale  to set the glueball mass. For very small quark mass the glueball becomes the heaviest state in both cases. The gauge coupling is sufficiently strong above the quark mass scale that the BF bound is violated and the quarks acquire a dynamical mass. The pure glue running between that scale and the IR pole is very fast since we are already at strong coupling when the quarks decouple - the glueball mass is set by essentially the quark decoupling scale. The interesting difference between the two cases with different $N_f$ is in the intermediate regime. The crossover between these two cases is fast for the $N_f=3$ theory but much slower for the walking $N_f=11$ theory. The reason is that for a range of intermediate quark mass scales the walking theory has run to a strong regime but which is below the critical coupling to trigger chiral symmetry breaking. Since it is walking, this regime, in which the quark decoupling and IR pole values are reasonably close, is enlarged in the walking theory - the cross over occurs over a wider range of quark mass. This is a signal in the spectra of walking behaviour. Such a signal is important because it does not depend on gauge dependent objects such as the coupling itself.

\section{Summary}

In this paper we have studied the lowest lying meson spectra ($\rho,\pi,\sigma$ and the lightest glueball) in asymptotically free gauge theory using some simple holographic models. In section 2 we studied an old top-down model \cite{Babington:2003vm} that had been previously shown \cite{ERev} to reproduce quenched lattice results in the $M_\rho-M_\pi$ plane (see Fig \ref{CMvslat}). The model includes strongly coupled gauge fields but returns to ${\cal N}=4$ Super Yang Mills in the UV. We argued, following \cite{Alvares:2012kr}, that the mesonic sector of the holographic model is described by a DBI action and that the only input of the background geometry is the running anomalous dimension of the quark bilinear operator, $\gamma$. We have shown that that function is a good approximation to quenched QCD in Fig \ref{CMvsQCD} and this explains the success of the model in the mesonic sector.

Motivated by this observation (and the inability to find a true gravity dual of the theories under analysis) we turned to the phenomenological Dynamic AdS/QCD model \cite{Alho:2013dka}. The model is basically the DBI action of a probe brane with the gauge dynamics input through a running scalar mass dual to $\gamma$. Once the running of $\gamma$ is input the spectrum is then predicted. Of course we do not know the non-perturbative running of $\gamma$, but the perturbative two loop running for the gauge coupling $\alpha$ combined with the one loop expression for $\gamma$ provide a well explored proposal - these theories lose asymptotic freedom for large quark numbers and have a conformal window with a fixed point for $\gamma$ at lower $N_f$, which rises as $N_f$ shrinks. In the holographic setting the condition $\gamma=1$ corresponds to a violation of the BF bound in AdS and the quark condensate switches on at lower values of $N_f$. The model then allows us to study any asymptotically free gauge theory with quarks in arbitrary representation and for any $N_c$ and $N_f$ - these quantities simply enter through the running of $\gamma$.

We first explored theories close to QCD - theories with a small number of quark flavours that do not impact on the running of the gauge coupling too much. These theories are characterized by a scale where $\gamma=1$ and chiral symmetry breaking is  triggered and a fast running of $\gamma$ at that scale. They all predict a very similar mesonic spectrum (see Fig \ref{CMvslat} for example) that match quenched lattice predictions \cite{Bali:2013kia} and the top-down model previously studied, at the level of errors in the lattice computations. A variant spectrum can be seen if $N_f$ is increased at fixed $N_c$, so that the theory approaches the edge of the conformal window and the running of $\gamma$ is slow when $\gamma=1$ (see for example the behaviour in SU(3) gauge theory in Fig \ref{su3fund}). The deviations are those expected given that the UV quark condensate is enhanced in these walking theories. Thus the holographic duals of walking models lead to significant
deviations in the `Edinburgh' meson plots, depending on the slope of the
anomalous dimension in the IR near $\gamma=1$. However, in the QCD-like
theories discussed above, these deviations are much smaller. This is due
to the fact that the QCD-like models are close to the quenched limit, in
which the derivative of $\gamma$ has less $N_f$ dependence at the chiral
symmetry breaking point. The result that the deviations are much smaller
in the QCD-like models as compared to the walking models adds further
weight to the success of the QCD-like models in matching lattice data.

The deviations in the spectrum due to walking is the only gross feature we have found across a range of gauge theories studied. For those theories we see the $\rho$ mass enhanced, the $\pi$ mass enhanced to a lesser degree at a fixed bare quark mass and the $\sigma$ mass falls as it acts as a Goldstone of the shift symmetry in the flattening effective potential for $\bar{q}q$. We show this behaviour in Fig \ref{sigmasu3fund} for SU(3) gauge theory. We explored theories with quarks in the adjoint and two index symmetric and anti-symmetric representations also and found similar behaviours. 

Finally we estimated the lightest glueball mass in the theories, although our model does not include mixing effects with the mesonic states. At large bare quark mass the quarks decouple at scales exponentially separated from the scale where the glue becomes strongly coupled and the glueball is light relative to the mesons. At zero quark mass, guided by QCD, the glueballs are expected to be heavier than the lightest mesons. We sketch the crossover behaviour with the quark mass for SU(3) gauge theory with $N_f=3,11$ in Fig \ref{glu} - the effect of walking dynamics in the physical spectra is to enlarge the energy range over which this cross-over occurs. This is a potententially useful statement of a signal for walking dynamics that is not couched in terms of a gauge dependent object such as the running coupling.

\bigskip
\bigskip

\noindent {\bf Acknowledgements:} We thank Biagio Lucini for discussions. NE and MS's work is supported by STFC.

\end{document}